\documentclass{webofc}
\usepackage[varg]{txfonts}   
\newcommand{\Ac}[1]{#1}
\newcommand{\Bc}[1]{#1}
\newcommand{\mbf}{\mathbf}

\begin{document}
\title{Studying the 3+1D structure of the Glasma using the weak field approximation}
%
%

\author{\firstname{Andreas} \lastname{Ipp}\inst{1}\fnsep\thanks{\email{ipp@hep.itp.tuwien.ac.at}} \and
        \firstname{Markus} \lastname{Leuthner}\inst{1}\fnsep\thanks{\email{markus.leuthner@tuwien.ac.at}} \and
        \firstname{David I.} \lastname{M\"uller}\inst{1}\fnsep\thanks{Presenter and corresponding author, \email{dmueller@hep.itp.tuwien.ac.at}} \and
        \firstname{Soeren} \lastname{Schlichting}\inst{2}\fnsep\thanks{\email{sschlichting@physik.uni-bielefeld.de}} \and
        \firstname{Pragya} \lastname{Singh}\inst{3, 4}\fnsep\thanks{\email{prasingh@jyu.fi}}
}

\institute{Institute for Theoretical Physics, TU Wien, A-1040 Vienna, Austria 
\and
           Fakult\"at f\"ur Physik, Universit\"at Bielefeld, D-33615 Bielefeld, Germany
\and
           Department of Physics, University of Jyv\"askyl\"a, FI-40014 Jyv\"askyl\"a, Finland
           \and
           Helsinki Institute of Physics, University of Helsinki, FI-00014 Helsinki, Finland
          }

\abstract{
We extend the weak field approximation for the Glasma beyond the boost-invariant approximation, which allows us to compute rapidity-dependent observables in the early stages of heavy-ion collisions. We show that in the limit of small fields, the weak field approximation agrees quantitatively with non-perturbative lattice simulations. Furthermore, we demonstrate that the rapidity profile of the transverse pressure is determined by longitudinal color correlations within the colliding nuclei.
}
\maketitle
\section{Introduction}
\label{intro}
The Color Glass Condensate (CGC) \cite{Gelis:2010nm} is an effective theory for high energy QCD. According to this framework, the pre-equilibrium medium created in heavy-ion collisions is dominated by classical color fields known as the Glasma \cite{Lappi:2006fp}. At leading order, the dynamics of the Glasma are determined by the classical Yang-Mills equations. Due to their non-linear nature, there are no general analytical solutions to the field equations and thus, approximations and simplifications are necessary. At very high collision energies, the boost invariant approximation, namely assuming rapidity independence of observables, allows us to simplify the problem from 3+1 dimensions to effectively 2+1 dimensions. Even the most state-of-the-art simulations of the early stages such as IP-Glasma \cite{Schenke:2012hg, Schenke:2012wb} rely on boost invariance. Though this approximation is only strictly valid in the limit of infinite collision energies, it is sufficient for observables at mid-rapidity. In addition, boost-invariant Glasma simulations can be carried out with initial conditions provided by the JIMWLK evolution \cite{Schenke:2016ksl, Schenke:2022mjv}, which re-introduces a dependence on rapidity, but the full details about the longitudinal structure and dynamics of high energy nuclei are at least partially lost. To remedy this problem, 3+1 dimensional lattice simulations of the Glasma have been carried out in the past \cite{Ipp:2020igo, Schlichting:2020wrv}. A drawback of this approach is that accurate simulations typically require large lattice sizes and are limited by computational resources. 

In these proceedings, we present a novel alternative approach to the 3+1 dimensional dynamics of the Glasma based on an extension of the weak field approximation \cite{Ipp:2021lwz}, which we show to be in quantitative agreement with non-perturbative simulations in the limit of weak fields, and, due to severely reduced computational effort, allows us to describe the genuinely three-dimensional dynamics of the Glasma for much larger systems than previously accessible through lattice simulations. In particular, we are able to compute the rapidity dependence of the Glasma, which arises due to correlations within the finite longitudinal extent of the colliding nuclei.

\section{Color Glass Condensate and the Glasma}

Within the CGC framework, valid at very high energies, the two colliding nuclei ``$A$'' and ``$B$'' can be described as highly Lorentz contracted sheets of colored glass, where hard partons (large Bjorken $x$) are approximated as classical color currents $\mathcal{J}^\mu$, which act as sources for a classical color field $\mathcal{A}^\mu$ describing the soft (small $x$) gluonic content of the nuclei. Adopting light-cone coordinates $x^\pm = (t \pm z) / \sqrt{2}$ and transverse coordinates $\mathbf x_\perp = (x,y)$, the light-like color currents of the nuclei can be written as $\mathcal J^\mu_{A,B}(x^\pm, \mathbf x_\perp) = \delta^{\mu\mp} \rho_{A,B}(x^\pm, \mathbf x_\perp)$, where $\rho_{A,B}$ are the color charge densities. Due to Lorentz contraction, the support along the light-like coordinates $x^\pm$ is assumed to be much smaller than in the transverse plane spanned by $\mathbf x_\perp$. Since the precise details of the colliding nuclei are a priori unknown, the color charge densities are treated as stochastic variables distributed according to probability functionals $W_{A,B}[\rho_{A,B}]$, which have to be specified by nuclear models. At leading order in the strong coupling constant $g$, the associated soft color fields $\mathcal A^\mu_{A,B}$ are determined by the Yang-Mills equations
\begin{align} \label{eq:ym_cgc}
    \mathcal D_\mu \mathcal{F}^{\mu\nu}_{A,B} = \mathcal{J}^\nu_{A,B}, \quad \mathcal{D}_\mu \mathcal{J}^\mu_{A,B} = 0,
\end{align}
where the field strength tensor is given by
\begin{align}
    \mathcal F^{\mu\nu}_{A,B} = \partial^\mu \mathcal{A}^\nu_{A,B} - \partial^\nu \mathcal{A}^\mu_{A,B} - ig \left[\mathcal{A}^\mu_{A,B}, \mathcal{A}^\nu_{A,B}\right],
\end{align}
and the gauge covariant derivative is $\mathcal D_\mu (\dots) = \partial_\mu - i g \left[ \mathcal{A}_\mu, \dots \right]$.
In covariant gauge $\partial_\mu \mathcal{A}^\mu = 0$, the Yang-Mills equations are solved by
\begin{align} \label{eq:sol_cgc}
    \mathcal{A}^\mp_{A,B}(x^\pm, \mathbf x_\perp) = \intop \frac{d^2 k_\perp}{(2\pi)^2} \frac{\tilde{ \mathcal{J}}^\mp_{A,B}(x^\pm, \mathbf k_\perp)}{\mathbf k_\perp^2 + m^2} e^{-i \mathbf k_\perp \cdot \mathbf x_\perp}, \quad \mathcal{A}^\pm_{A,B} = \mathcal{A}^i_{A,B} = 0,
\end{align}
where $m$ is an infrared regulator (on the order of the confinement scale), which suppresses infrared modes, and $\tilde{ \mathcal{J}}^\mp_{A,B}(x^\pm, \mathbf k_\perp)$ are the Fourier transformed currents.

\begin{figure}[t]
\sidecaption
\centering
\includegraphics[width=0.4\textwidth]{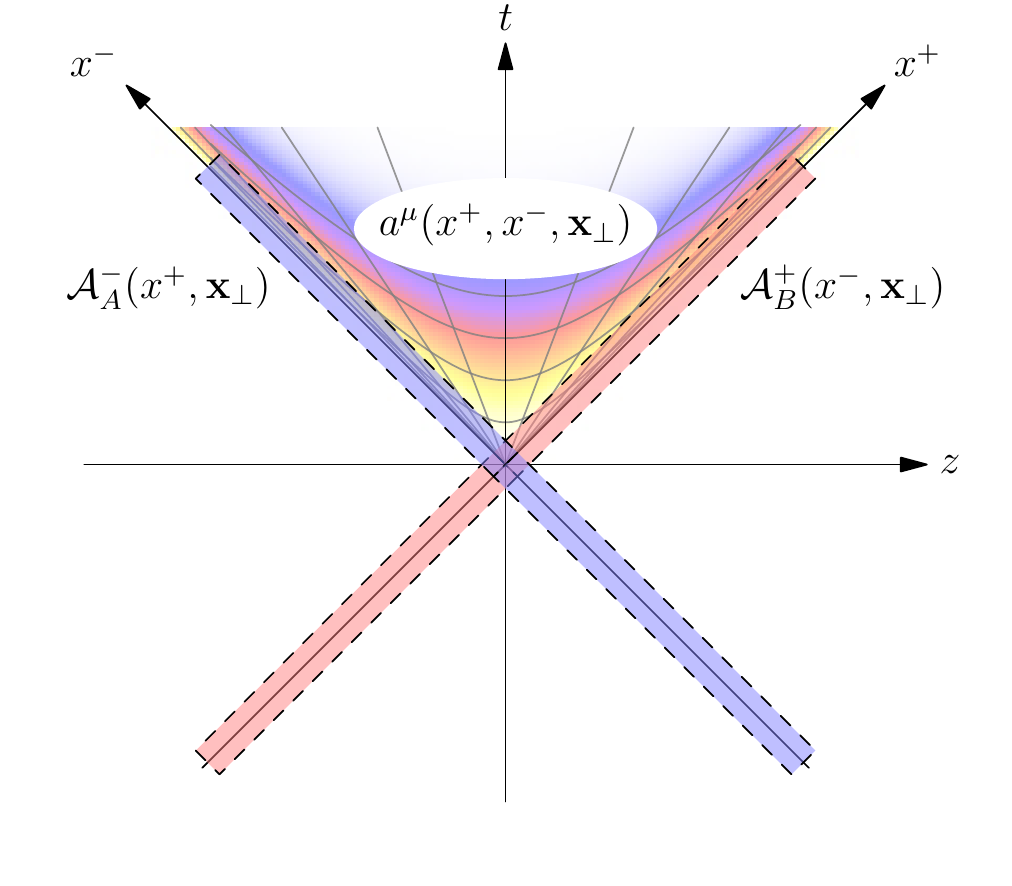}
\caption{Minkowski diagram of the 3+1 dimensional collision problem. The background color fields and color currents of the nuclei $\mathcal{A}^\mu_{A,B}$ and $\mathcal{J}^\mu_{A,B}$ are strongly peaked around the light cone. The thin, but finite support along the longitudinal directions is indicated by red and blue rectangles. The color field of the Glasma $a^\mu$  in the future light cone is produced due to the non-linear interactions of the high energy nuclei in the collision center.}
\label{fig:collision_problem}       
\end{figure}

The classical color field produced in the collision of two CGCs is determined by solving the collision problem
\begin{align} \label{eq:ym_glasma}
    D_\mu {F}^{\mu\nu}(x) = J^\nu(x), \quad D_\mu J^\mu(x) = 0,
\end{align}
with
\begin{align}
     F^{\mu\nu}= \partial^\mu {A}^\nu - \partial^\nu {A}^\mu - ig \left[{A}^\mu, {A}^\nu\right].
\end{align}
and initial conditions provided in the asymptotic past
\begin{align}
    \lim_{t \rightarrow -\infty} J^\nu(x) &= \mathcal{J}^\nu_{A}(x^+, \mathbf x_\perp) + \mathcal{J}^\nu_{B}(x^-, \mathbf x_\perp), \\
    \lim_{t \rightarrow -\infty} A^\nu(x) &= \mathcal{A}^\nu_{A}(x^+, \mathbf x_\perp) + \mathcal{A}^\nu_{B}(x^-, \mathbf x_\perp),
\end{align}
which, in contrast to Eq.~\eqref{eq:ym_cgc}, has no general closed form solution due to the non-linear nature of the Yang-Mills equations. We use calligraphic symbols to denote the freely propagating (non-interacting) solutions, while the non-calligraphic symbols denote the full solution. A Minkowski diagram of the collision is shown in Fig.~\ref{fig:collision_problem}, where the Glasma is the color field in the future light cone of the collision. Over the past decades, a number of methods and approximations have been devised to obtain approximate solutions to Eq.~\eqref{eq:ym_glasma}, the most notable of which is the boost-invariant approximation or the ultrarelativistic limit \cite{Kovner:1995ja}. In this limit, the nuclei are assumed to be contracted to infinitesimally thin disks, such that the color currents reduce to $\mathcal J^\mu_{A,B}(x^\pm, \mathbf x_\perp) = \delta^{\mu\mp} \delta(x^\pm) \rho_{A,B}(\mathbf x_\perp)$. Due to invariance under longitudinal boosts, the collision problem can be treated as effectively 2+1 dimensional, since gauge-invariant observables $\mathcal{O}(\tau, \mathbf x_\perp)$ in the future light cone must be independent of space-time rapidity $\eta = \frac{1}{2} \ln (x^+ / x^-)$ and only depend on proper time $\tau = \sqrt{2x^+ x^-}$ and transverse coordinates $\mathbf x_\perp$. The boost-invariant approximation enables the non-perturbative determination of the produced color field on the inside of the future light-cone, which can be taken as initial conditions at $\tau \rightarrow 0^+$ for a source-free evolution
\begin{align}
      D_\mu {F}^{\mu\nu}(\tau, \mathbf x_\perp) = 0, \qquad {A}^\mu(\tau \rightarrow 0^+, \mathbf x_\perp) =  A^\mu_\mathrm{Glasma}(\mathbf x_\perp),    
\end{align}
where the fields $A^\mu_\mathrm{Glasma}(\mathbf x_\perp)$ are provided by the  Glasma initial conditions \cite{Kovner:1995ja}. To carry out the time evolution for $\tau > 0$, the 2+1 dimensional Yang-Mills equations can be solved using classical lattice methods \cite{Kovner:1995ja, Krasnitz:1998ns, Lappi:2003bi}, a Taylor series in $\tau$ \cite{Chen:2015wia}, or, in the case of small color currents $\mathcal{J}^\mu_{A,B}$, using a weak-field approximation \cite{Guerrero-Rodriguez:2021ask}. However, in order to go beyond boost invariance, the Glasma initial conditions are insufficient as they strictly rely on the assumption that the nuclei are infinitesimally thin along the light-like coordinates. As a brute-force approach, one may solve the full problem in Eq.~\eqref{eq:ym_glasma} using 3+1 dimensional lattice simulations either based on the colored particle-in-cell method \cite{Ipp:2020igo} or dynamical currents \cite{Schlichting:2020wrv}. In both cases, large lattices are required to obtain accurate numerical approximations to the Glasma. As a semi-analytic alternative to these purely numerical lattice approaches, we explore how to extend the weak-field approximation to 3+1 dimensions.

\section{Weak field approximation in 3+1 dimensions}
\label{weak}

The weak field approximation is a perturbative method to solve the Yang-Mills equations \eqref{eq:ym_glasma} which relies on an expansion in terms of the color currents $\mathcal{J}^\mu_{A,B}$. Details about this calculation can be found in our paper \cite{Ipp:2021lwz}. The idea is to perform a split of the gauge fields $A^\mu$ and currents $J^\mu$ into background currents and fields provided by Eq.~\eqref{eq:sol_cgc}, and perturbations $j^\mu$ and $a^\mu$
\begin{align}
    A^\mu(x) &= \mathcal{A}^\mu_A(x) + \mathcal{A}^\mu_B(x) + a^\mu(x), \label{split} \\
    J^\mu(x) &= \mathcal{J}^\mu_A(x) + \mathcal{J}^\mu_B(x) + j^\mu(x),
\end{align}
which we take as an Ansatz to solve Eq.~\eqref{eq:ym_glasma}. Adopting covariant gauge and expanding the Yang-Mills equations to leading order in the background currents (the dilute limit), yields a 3+1 dimensional wave equation for the perturbative gauge field
\begin{align}
    \partial_\nu \partial^\nu a^\mu(x) = \mathcal{S}^\mu[\mathcal{J}_A, \mathcal{J}_B] = O(\mathcal{J}_A \mathcal{J}_B),
\end{align}
where $\mathcal{S}^\mu$ is a functional of the background currents and acts as a source term for $a^\mu$. The wave equation can be solved using the method of Green's functions and its solution can be further simplified by exploiting the space-time structure of the color currents.

Having obtained an explicit solution for $a^\mu$, the perturbative field strength tensor ${f^{\mu\nu} = \partial^\mu a^\nu - \partial^\nu a^\mu}$ assumes a particularly simple form given by
\begin{align} 
    f^{+-}(x) &= - \frac{g}{2\pi} \intop d^2 \mbf u_\perp \intop_{-\infty}^{+\infty} d\eta_z \, V(x, \mathbf u_\perp, \eta_z), \label{eq:f_pm} \\ 
    f^{\pm i}(x) &= +\frac{g}{2\pi} \intop d^2 \mbf u_\perp  \intop_{-\infty}^{+\infty} d\eta_z \, \big(V^{ij}(x, \mathbf u_\perp, \eta_z) \, \mp \, \delta^{ij} V(x, \mathbf u_\perp, \eta_z)\big) \frac{w^j}{\sqrt{2}} e^{\pm \eta_z}, \label{f_pmi} \\ 
    f^{ij}(x) &= - \frac{g}{2\pi} \intop d^2 \mbf u_\perp  \intop_{-\infty}^{+\infty} d\eta_z \, V^{ij}(x, \mathbf u_\perp, \eta_z), \label{eq:f_ij}
\end{align}
where the unit vector $\mathbf w$ is given by $w^i = (x^i_\perp - u^i_\perp) / | \mathbf x_\perp - \mathbf u_\perp|$ and
\begin{align}
    V(x, \mathbf u_\perp, \eta_z) &= f_{abc} t^c \,
    \Ac{\partial^i \mathcal{A}^{a, -}_A(x^+ - {\small \frac{|\mbf x_\perp - \mbf u_\perp|}{\sqrt 2}} e^{+\eta_z}, \mbf u_\perp)} \,\,    \Bc{\partial^i \mathcal{A}^{b, +}_B(x^- - {\small \frac{|\mbf x_\perp - \mbf u_\perp|}{\sqrt 2}} e^{-\eta_z}, \mbf u_\perp)}, \\
    V^{ij}(x, \mathbf u_\perp, \eta_z) &= f_{abc} t^c \, \big(     \Ac{\partial^i \mathcal{A}^{a, -}_A(\dots)}    \, \Bc{\partial^j \mathcal{A}^{b, +}_B(\dots)}     - \Ac{\partial^j \mathcal{A}^{a, -}_A(\dots)}     \,\Bc{\partial^i \mathcal{A}^{b, +}_B(\dots)}     \big).
\end{align}
Here, $t^a$ are the Hermitian generators of SU($N_c$) and $f_{abc}$ are the associated structure constants. Given background fields $\mathcal{A}^\mu_{A,B}$, the perturbative field strength tensor can be obtained by numerical integration. 
Using the field strength tensor, the perturbative energy-momentum tensor of the Glasma can be immediately computed from
\begin{align}
t^{\mu\nu}(x) = 2 \mathrm{Tr} \big[f^{\mu \rho}(x) f_\rho{}^\nu(x) + \frac{1}{4} g^{\mu\nu} f^{\rho\sigma}(x) f_{\rho\sigma}(x)\big],
\end{align}
which, in particular, allows us to determine the energy density and the pressure components.

\section{A 3+1 dimensional McLerran-Venugopalan model} \label{nuclear_model}

\begin{figure*}
\centering
\includegraphics[width=0.32\textwidth]{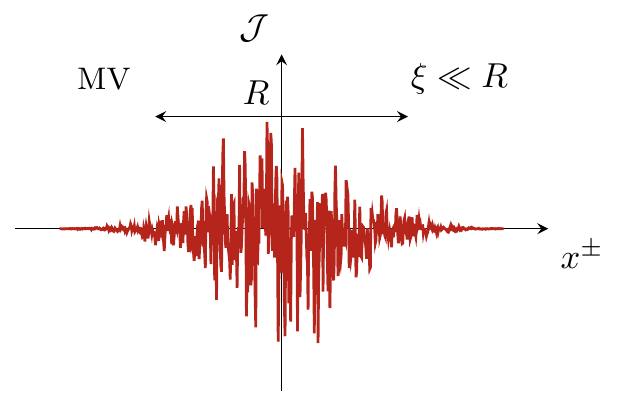}
\includegraphics[width=0.32\textwidth]{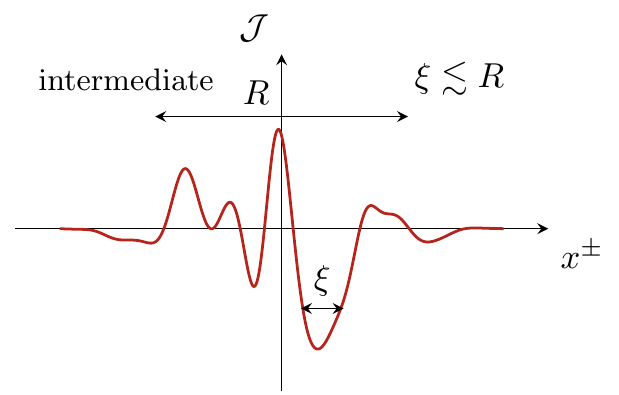}
\includegraphics[width=0.32\textwidth]{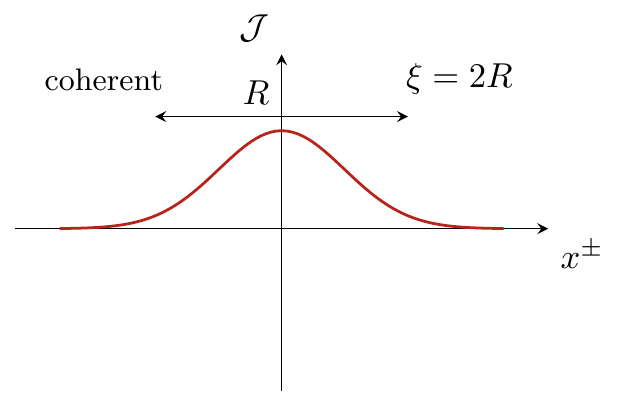}
\caption{A sketch of the color current in various limits of the 3+1 dimensional MV model. The ratio of the longitudinal width $R$ and the longitudinal correlation length $\xi$ determines the non-trivial structure of the nuclei. For $\xi / R \rightarrow 0$ (left panel) the correlator reduces to the usual MV model. The coherent limit is reached by choosing the correlation length to be as large as the system ($\xi=2R$, right panel). The longitudinal structure of realistic high energy nuclei is most likely described by $\xi \lesssim R$ for not too small $\xi$ (center panel).}
\label{fig:nuclear_model}       
\end{figure*}

Our main result in Eqs.~\eqref{eq:f_pm}--\eqref{eq:f_ij} holds for any choice of the color currents $\mathcal{J}_A(x^+, \mathbf x_\perp)$ and $\mathcal{J}_B(x^+, \mathbf x_\perp)$ at leading order. However, to compute observables such as $t^{\mu\nu}$, a specific nuclear model has to be chosen by specifying the probability functionals $W_{A,B}[\rho_{A,B}]$. In \cite{Ipp:2021lwz} we used a 3+1 dimensional extension of the originally 2+1 dimensional McLerran-Venugopalan (MV) model \cite{McLerran:1993ni, McLerran:1993ka}, which introduces non-trivial longitudinal color structure along the light-like directions $x^\pm$. Since the MV model is Gaussian, it is sufficient to specify the one- and two-point functions
\begin{align}
\langle \rho^a_{A,B}(x^\pm, \mathbf{x_\perp}) \rangle &= 0, \label{eq:onep}\\
\langle \rho^a_{A,B}(x^\pm, \mathbf{x_\perp}) \rho^b_{A,B}(x'^\pm, \mathbf x'_\perp) \rangle &= g^2 \mu^2_{A,B}\delta^{ab} T_R(\frac{x^\pm + x'^\pm}{2}) U_\xi(x^\pm - x'^\pm)\delta^{(2)}(\mathbf x_\perp- \mathbf x'_\perp). \label{eq:twop}
\end{align}
The strength of the color charge fluctuations is determined by the parameter $g^2 \mu^2_{A,B}$ (in units of energy squared). The longitudinal structure is determined by the two functions $T_R$ and $U_\xi$, which we choose to be normalized Gaussians of width $R$ and $\xi$. The longitudinal length scale $R$ is related to the Lorentz-contracted extent of the nuclei and $\xi \leq 2R$ determines the longitudinal correlation length within a nucleus. A sketch for various choices of $\xi/R$ is shown in Fig.~\ref{fig:nuclear_model}. In the limit of $\xi \rightarrow 0$ we recover the original MV model, whereas the limit of maximal correlation $\xi \rightarrow 2R$ corresponds to nuclei with coherent color structure along $x^\pm$ (the coherent limit).

\section{Numerical results} 
\label{results}

\begin{figure*}[t]
\centering
\includegraphics[width=\textwidth]{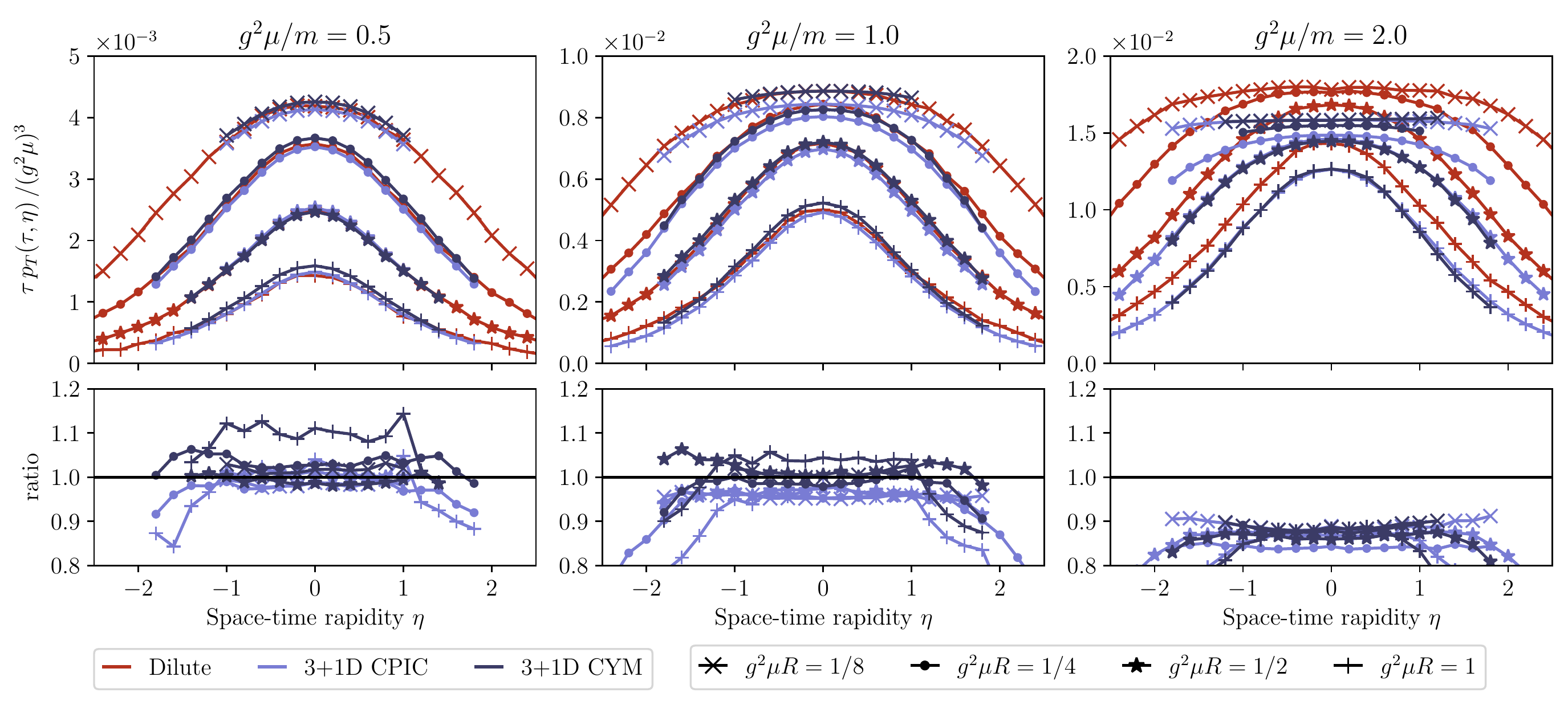}
\caption{Rapidity profiles of transverse pressure as a function of space-time rapidity $\eta$ in the coherent limit $\xi = 2R$ for various values of the longitudinal width $R$ and the non-linearity parameter $g^2\mu/m$ at late times $\tau \gg R$. The red line corresponds to the weak field approximation, whereas the two purple lines correspond to two different lattice implementations (3+1D colored particle-in-cell (CPIC) \cite{Ipp:2020igo} and 3+1D classical Yang-Mills (CYM) using dynamical currents \cite{Schlichting:2020wrv}). Both the magnitude and the rapidity dependence are well described by the weak field approximation for small fields $g^2 \mu / m \lesssim 1$ (left and center panels), but the weak-field pressure is generally overestimated in the more non-linear case (right panel). Nevertheless, the rapidity profile is still correctly reproduced for larger $g^2\mu /m$ as can be seen from the bottom panels, where we show the ratio of the weak field approximation to the lattice simulations. Adapted from \cite{Ipp:2021lwz}.}
\label{fig:comparison}       
\end{figure*}

Equipped with a nuclear model and having worked out the analytical result for the field-strength tensor, we can now compute observables in the 3+1 dimensional Glasma. Expectation values of observables such as $t^{\mu\nu}$ can be computed in at least two different ways: either event-by-event by numerically integrating Eqs.~\eqref{eq:f_pm}--\eqref{eq:f_ij} for each collision event separately, or alternatively, by inserting the one- and two-point functions Eqs.~\eqref{eq:onep} and \eqref{eq:twop} into $t^{\mu\nu}$, which takes care of the event average. In our paper \cite{Ipp:2021lwz} we chose the latter approach and employed Monte Carlo sampling to approximate the integral.

\subsection{Comparison to non-perturbative lattice simulations} \label{comparison}

We first verify that our calculation using the weak field approximation agrees with non-perturbative 3+1 dimensional lattice simulations in the dilute limit. We focus on the transverse pressure of the Glasma 
\begin{align}
    p_T(x) = \langle t^{xx} (x) \rangle = \langle t^{yy} (x) \rangle,
\end{align}
which is solely due to longitudinal color-electric and -magnetic fields of the Glasma and allows for a clean comparison of the analytical and lattice calculations. Furthermore, we focus on the coherent limit $\xi = 2R$ as this particular case requires less resolution along the longitudinal axis to properly resolve the longitudinal structure and can thus be simulated on the lattice with sufficient numerical accuracy. Numerical results are shown in Fig.~\ref{fig:comparison}, where we plot the transverse pressure as a function of space-time rapidity $\eta$ at fixed proper time $\tau \gg R$ for various values of $R$ and $g^2 \mu / m$. The latter parameter controls the non-linearity of the model with the dilute limit being realized for $g^2 \mu / m \ll 1$. We observe that the weak field results agree  remarkably well with the lattice up to $g^2 \mu / m \simeq 1$. For larger values, the weak field result overestimates the overall pressure, but is still able to accurately reproduce the rapidity profile of the Glasma. 

\subsection{Longitudinal correlations and rapidity-dependent transverse pressure} \label{correlations}

\begin{figure}[t]
\sidecaption
\centering
\includegraphics[width=0.6\textwidth]{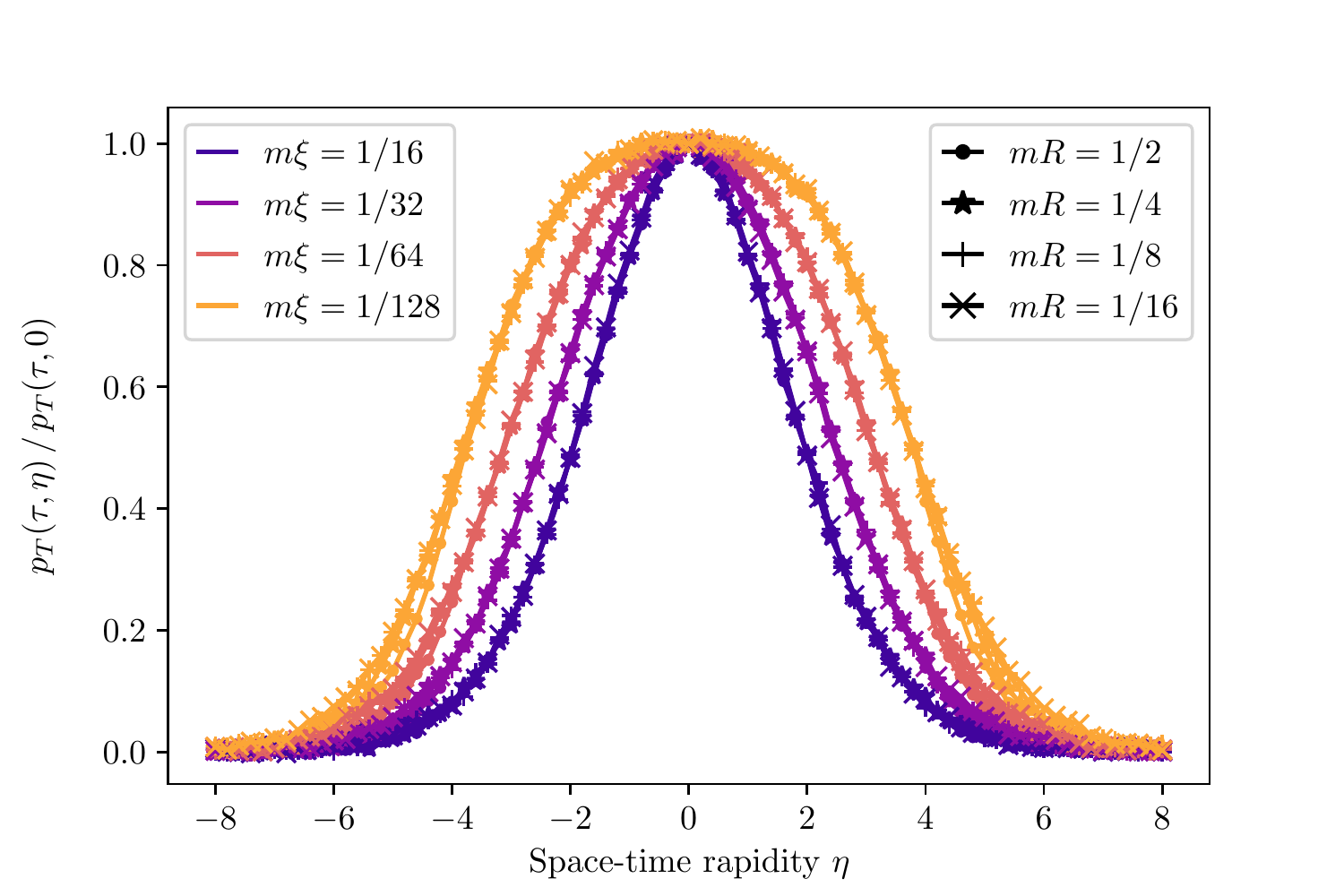}
\caption{Normalized rapidity profile of the transverse pressure $p_T$ as a function of space-time rapidity $\eta$ for various values of $R$ and $\xi$ ($\xi \lesssim R$) at late times $\tau \gg R$. We observe that the width of the profile only weakly depends on the dimensionless longitudinal width $mR$ (same color, various markers), but strongly depends on the longitudinal correlation length $m\xi$ (various colors). }
\label{fig:rapidity_and_correlations}       
\end{figure}

Having demonstrated the correctness of the weak field approach, we can now study the dependence of the rapidity profile in the 3+1 dimensional Glasma on the longitudinal parameters $R$ and $\xi$, in particular in the intermediate regime $\xi \lesssim R$. A plot of the transverse pressure rapidity profile for various values of $R$ and $\xi$ is shown in Fig.~\ref{fig:rapidity_and_correlations}. Remarkably, we find that the width of the rapidity profile depends only weakly on the longitudinal system size $R$. Instead, the profile width is mostly determined by the size of the longitudinal correlations $\xi$. This demonstrates that, at least in the dilute limit, the rapidity dependence of the initial energy density of the early stages can be directly traced back to the non-trivial longitudinal structure of the colliding nuclei. 

\section{Conclusion} \label{conclusion}

The weak field approximation is a perturbative approach to compute observables in the Glasma in high energy heavy-ion collisions based on linearizing the Yang-Mills equations around non-interacting background fields and color currents. We have demonstrated that this method can be extended beyond the 2+1 dimensional boost-invariant approximation, which allows us to access the 3+1 dimensional space-time structure of the Glasma in a semi-analytical manner. In particular, it allows us to probe the Glasma at parameters that are typically difficult to achieve in lattice simulations due to high computational cost, namely the case of intermediate correlation length $\xi \lesssim R$. We have shown that the weak field approximation agrees with non-perturbative lattice simulations in the dilute limit for coherent nuclei. Going towards the limit of smaller correlations lengths, we observed that the rapidity distribution and thus the longitudinal dynamics of the Glasma are not sensitive to the size of the system. Our calculations shows that the widths of the rapidity profiles are mostly due to the longitudinal color structure within the colliding nuclei. It should be noted that the numerical results in these proceedings are based on a simple extension of the MV model without transverse structure. Going forward, we plan to study the 3+1 dimensional Glasma in the dilute limit for more realistic systems by including realistic collision geometry and by accounting for nucleonic and sub-nucleonic structure using hot spots \cite{Demirci:2021kya, Demirci:2022wuy, Mantysaari:2022ffw}.

\vspace{1em}

\begin{acknowledgement}
We thank B.~Schenke and T.~Lappi for discussions and collaboration on related projects.
DM and ML are supported by the Austrian Science Fund FWF No.~P34764. ML further acknowledges funding from the Doktoratskolleg Particles and Interactions (DK-PI,  FWF doctoral program No.~W-1252-N27). SS and PS are supported under the Deutsche Forschungsgemeinschaft (DFG, German Research Foundation) through the CRC-TR 211 `Strong-interaction matter under extreme conditions' - project number: 315477589 TRR-211. PS is also supported by the
Academy of Finland, project 321840 and under the European Union’s Horizon 2020 research and innovation programme by the STRONG-2020 project (grant agreement
No 824093). The computations in this work were performed at the Paderborn Center for Parallel Computing (PC2) and the Vienna Scientific Cluster (VSC).
\end{acknowledgement}

\bibliography{references}

\end{document}